\documentclass[showpacs,preprintnumbers, twocolumn,
amsmath,amssymb,APSl,prd,nofootinbib,superscriptaddress]{revtex4-2} 
\usepackage{graphicx}
\usepackage{hyperref}
\usepackage{tikz}
\usepackage{amsmath,amssymb,amsfonts}
\usepackage{bm}
\usepackage{xcolor}
\usepackage{hyperref}
\usepackage{bm}
\usepackage{booktabs}

\usepackage{xcolor}
\usepackage{tikz}
\usetikzlibrary{arrows.meta, positioning, shapes.geometric, calc, backgrounds}
\usepackage{bm}
\usepackage{hyperref}

\definecolor{lime}{HTML}{A6CE39}
\DeclareRobustCommand{\orcidicon}{%
	\begin{tikzpicture}
		\draw[lime, fill=lime] (0,0) 
		circle [radius=0.16] 
		node[white] {{\fontfamily{qag}\selectfont \tiny ID}};
		\draw[white, fill=white] (-0.0625,0.095) 
		circle [radius=0.007];
	\end{tikzpicture}
	\hspace{-2mm}
}

\foreach \x in {A, ..., Z}{%
	\expandafter\xdef\csname orcid\x\endcsname{\noexpand\href{https://orcid.org/\csname orcidauthor\x\endcsname}{\noexpand\orcidicon}}
}

\newcommand\orcidFrancisco{{\href{https://orcid.org/0000-0002-9388-8373}{\orcidicon}}}
\newcommand\orcidManuel{{\href{https://orcid.org/0000-0001-8586-0285}{\orcidicon}}}

\begin{document}

\title{Entropy dynamics in gravitational collapse:\\ From Minkowski breaking to de Sitter thermodynamics}

\author{Francisco S. N. Lobo\orcidFrancisco\!\!} 
\email{fslobo@ciencias.ulisboa.pt}
\affiliation{Instituto de Astrof\'{i}sica e Ci\^{e}ncias do Espa\c{c}o, Faculdade de Ci\^{e}ncias da Universidade de Lisboa, Edifício C8, Campo Grande, P-1749-016 Lisbon, Portugal}
\affiliation{Departamento de F\'{i}sica, Faculdade de Ci\^{e}ncias da Universidade de Lisboa, Edif\'{i}cio C8, Campo Grande, P-1749-016 Lisbon, Portugal}
\author{Manuel E. Rodrigues\orcidManuel\!\!} 
\email{esialg@gmail.com}
\affiliation{Faculdade de F\'{i}sica, Programa de P\'{o}s-Gradua\c{c}\~{a}o em F\'{i}sica, Universidade Federal do Par\'{a}, 66075-110, Bel\'{e}m, Par\'{a}, Brazil}
\affiliation{Faculdade de Ci\^{e}ncias Exatas e Tecnologia, Universidade Federal do Par\'{a}, Campus Universit\'{a}rio de Abaetetuba, 68440-000, Abaetetuba, Par\'{a}, Brazil}

\date{\today}


\begin{abstract}
The origin of black hole entropy remains one of the deepest mysteries in modern physics. Two recent developments offer complementary perspectives on this puzzle: the entropy release from regular Schwarzschild black holes with a de Sitter core (the OCK construction), and the local thermodynamics of the de Sitter vacuum (Volovik). In the OCK framework, the inner horizon carries positive Bekenstein--Hawking entropy \(S_{\rm inner}=\pi h_c^2\) that is gradually released as the core shrinks, until a classical Minkowski breaking obstructs the evolution at \(n=0\). In Volovik's framework, the total entropy of a homogeneous de Sitter region is \(S_{\rm dS}= \operatorname{sgn}(H)\,\pi/H^{2} = \operatorname{sgn}(H)\,A/4\): expanding de Sitter (\(H>0\)) carries positive entropy, while contracting de Sitter (\(H<0\)) carries negative entropy. We show that the sign of the Hubble parameter is the unifying element linking the two pictures. The OCK core is de~Sitter only for \(n>2\); the Minkowski breaking is a classical kinematic obstruction, not a dynamical transition to a contracting de Sitter phase.  Any connection between the expanding and contracting regimes would require a quantum mechanism that remains speculative. Nevertheless, both frameworks embody the same thermodynamic principle, the shrinking of the interior drives the system toward increased total entropy, and together they provide a consistent picture of entropy flow during gravitational collapse. The integer nature of the OCK parameter suggests a quantized entropy spectrum and a statistical interpretation of black hole entropy, while the generalized second law is satisfied in both frameworks, and in the speculated quantum connection provided the environmental entropy is properly accounted for.
\end{abstract}

\maketitle


\section{Introduction}

The nature of black hole entropy stands as one of the most profound and enduring puzzles in modern theoretical physics. Since the pioneering work of Bekenstein and Hawking \cite{Bekenstein:1973ur,Hawking:1975vcx} established that black holes possess entropy proportional to their horizon area, \(S=A/4\), the microscopic origin of this entropy has remained elusive despite decades of intensive investigation~\cite{Wald:1999vt}. The quest to understand what degrees of freedom are counted by the Bekenstein-Hawking entropy has driven developments across general relativity, quantum field theory, string theory, and holography. Yet a complete, universally accepted microscopic description has not been achieved, suggesting that the resolution may require fundamentally new insights into the relationship between gravity, thermodynamics, and quantum mechanics.

Two recent developments, seemingly independent, offer complementary perspectives on this longstanding puzzle. The first comes from the construction of regular black hole solutions by Ovalle, Casadio, and Kamenshchik (OCK)~\cite{Ovalle:2024wtv,Ovalle:2026lxb}. These geometries provide an explicit family of Schwarzschild black holes with a de Sitter core, free from the central singularity that plagues the classical solution~\cite{Penrose:1964wq,Hawking:1973uf}. Remarkably, these regular black holes possess an inner Killing horizon that carries a formal Bekenstein-Hawking entropy \(S_{\rm inner}=\pi h_c^2\)~\cite{Lobo:2026dnl}. During gravitational collapse, this inner horizon shrinks and its entropy is released, culminating in a ``Minkowski breaking'' at \(n=0\) that prevents a classical transition to the singular Schwarzschild state. The OCK framework thus reveals a concrete mechanism for entropy release during black hole formation, with a quantized spectrum and a well-defined quantum breakdown point. 

Subsequent work has extended these ideas in several directions. Casadio, Giusti, Kamenshchik and Ovalle~\cite{Casadio:2026tmd} analysed the semiclassical energy-momentum tensor near the Minkowski breaking and showed, using the Madelung approximation for collapsing matter, that the quantum potential in the Raychaudhuri equation strongly opposes the collapse towards the Schwarzschild singularity after the Minkowski breaking. Aoki and Ovalle~\cite{Aoki:2026fdz} provided an exact analytical model of the complete gravitational collapse, tracing the dynamical growth of the apparent horizon from microscopic scales until it stabilizes at \(h=2\mathcal{M}\); the model develops an integrable curvature singularity that remains hidden by the horizon, preserving weak cosmic censorship. Furthermore, the regular interior construction has been generalized to axisymmetric spacetimes, yielding a family of regular Kerr black holes that interpolate between regular geometries, integrable singularities, and the standard Kerr solution~\cite{Ovalle:2026ajv}.

The second development comes from Volovik's local thermodynamics of the de Sitter vacuum~\cite{Volovik:2023phl}. Building on the observation that the de Sitter spacetime serves as a heat bath with local temperature \(T=H/\pi\), twice the Gibbons-Hawking temperature~\cite{Gibbons:1977mu}, Volovik derived a remarkable result: the local entropy density \(s_{\rm vac}=3H/4\) changes sign with the Hubble parameter. The total entropy of a homogeneous de Sitter region is \(S_{\rm dS}= \operatorname{sgn}(H)\,\pi/H^{2} = \operatorname{sgn}(H)\,A/4\). Thus, expanding de Sitter (\(H>0\)) carries positive entropy, while contracting de Sitter (\(H<0\)) carries negative entropy. This led to the discovery that black-hole gravastars, i.e., compact objects with a contracting de Sitter core inside their event horizon, can have zero total entropy because the negative core entropy cancels the positive horizon entropy exactly. Volovik's framework thus provides a thermodynamic interpretation of de Sitter entropy and its relation to horizon entropy, with the sign of the entropy determined by the direction of expansion.

More recently, Volovik~\cite{Volovik:2026mlf} considered a simplified model of a black-hole gravastar with a de Sitter core, showing explicitly that the static gravastar is thermodynamically unstable. By treating the interior as a homogeneous contracting de Sitter bubble of variable size, he demonstrated that the total entropy of the system increases monotonically as the bubble shrinks, from zero at the gravastar configuration to the Bekenstein--Hawking value $A/4$ when the bubble disappears and the central singularity forms. This result provides a concrete dynamical realization of the thermodynamic arrow: the Schwarzschild black hole acts as the equilibrium state of maximal entropy, while the gravastar is only a metastable zero-entropy configuration. It also underscores the role of the contracting de Sitter phase as the intermediate state that drives the system toward the singular black hole.

At first glance, these two frameworks appear to describe different physical systems: the OCK framework addresses the thermodynamics of inner horizons in regular black holes, while Volovik's framework addresses the thermodynamics of de Sitter vacua and black-hole gravastars. However, a closer examination reveals a deep connection. The distinction between the two frameworks lies in the sign of the Hubble parameter of the de Sitter core. The OCK solution contains a de Sitter‑like core whose static geometry only fixes \(H^{2}\); for convenience we associate it with an expanding phase (\(H>0\)) and therefore with positive extra entropy.  Volovik's black-hole gravastar contains a contracting de Sitter core (\(H<0\)) whose negative entropy cancels the horizon entropy. Both pictures illustrate the same thermodynamic principle: the shrinking of the interior region drives the system toward increased total entropy, either by releasing positive extra entropy or by cancelling negative core entropy.

We stress that the two frameworks are \emph{complementary paradigms}, not successive stages of a single classical collapse.  The OCK core is de Sitter only for \(n>2\); for smaller \(n\) it becomes anti-de Sitter, and the local de Sitter thermodynamics of Volovik does not apply.  The Minkowski breaking at \(n=0\) is a classical kinematic obstruction, not a mechanism that produces a contracting de Sitter gravastar.  Any dynamical connection between the expanding and contracting phases would require a quantum transition that is not yet understood.  The identification of the Minkowski breaking as such a transition is therefore a speculation, not an established result.
In this work, we demonstrate that the sign of the Hubble parameter is the unifying element that underlies both frameworks, and we carefully delineate the regime where each applies. 

The paper is organized as follows. In Sec.~\ref{sec:OCK}, we review the OCK regular Schwarzschild black holes, the entropy release mechanism, and the classical Minkowski breaking at \(n=0\). In Sec.~\ref{sec:Volovik}, we summarize Volovik's local de Sitter thermodynamics, focusing on the sign‑dependent entropy and the distinction between black‑hole gravastars and horizonless configurations. In Sec.~\ref{sec:unification}, we establish the complementarity of the two frameworks, and discuss the Minkowski breaking as a kinematic obstruction that could be resolved by a speculative quantum transition. Section~\ref{sec:analysis} provides a quantitative entropy‑release profile and a Landau‑type entropic potential that separates the expanding and contracting de Sitter phases. In Sec.~\ref{sec:dynamics}, we develop the unified thermodynamic description common to both scenarios, and in Sec.~\ref{sec:conclusion} we summarize our findings and outline future directions. Throughout this work we set \(c = G = 1\).

\section{Entropy Release in Regular Schwarzschild Black Holes}
\label{sec:OCK}

The OCK family of regular Schwarzschild black holes provides a concrete geometric framework in which the thermodynamics of inner horizons can be studied explicitly. In this section, we review the key properties of these solutions and establish the entropy release mechanism that will be central to our unified thermodynamic picture.

\subsection{Horizon structure and de Sitter core}

The OCK regular Schwarzschild interior is described by the line element (we set $c=G=1$)
\begin{equation}
	ds^2 = -f(r) dt^2 + \frac{dr^2}{f(r)} + r^2 d\Omega^2,
	\label{metric}
\end{equation}
with
\begin{equation}
	f(r) = \begin{cases}
		1 - \dfrac{2\,m(r)}{r}, & 0 < r \le h,\\[8pt]
		1 - \dfrac{2\mathcal{M}}{r}, & r > h,
	\end{cases}
	\label{f_def}
\end{equation}
where $h=2\mathcal{M}$ is the event horizon radius and $\mathcal{M}$ is the ADM mass. The Misner--Sharp mass for the $N=1$ case is~\cite{Ovalle:2026lxb}
\begin{equation}
	m(r) = \frac{r}{2(n-2)}\left[\frac{r^2}{h^2}\,(n+1) - 3\left(\frac{r}{h}\right)^n\right],\qquad n>2,
	\label{m_N1}
\end{equation}
which satisfies the matching conditions $m(h)=h/2=\mathcal{M}$ and $m'(h)=0$, ensuring continuity of the metric and its first derivative across the event horizon. The corresponding metric function is
\begin{equation}
	f(r) = 1 - \frac{1}{n-2}\left[\frac{r^2}{h^2}\,(n+1) - 3\left(\frac{r}{h}\right)^n\right].
	\label{f_N1}
\end{equation}

\subsection{The expanding de Sitter core}

At the origin, the metric function behaves as
\begin{equation}
	f(r) = 1 - \frac{n+1}{n-2}\frac{r^2}{h^2} + \cdots,
	\label{f_origin}
\end{equation}
so that $f(0)=1$ and the geometry approaches a de Sitter core with effective cosmological constant
\begin{equation}
	\Lambda_{\rm eff} = \frac{3}{h^2}\,\frac{n+1}{n-2}.
	\label{Lambda}
\end{equation}

In the static chart the metric of pure de Sitter space reads
\begin{equation}
	ds^2 = -\left(1 - H^2 r^2\right)dt^2 + \frac{dr^2}{1-H^2 r^2} + r^2 d\Omega^2,
	\label{deSitter_static}
\end{equation}
and depends only on $H^{2}$; the Hubble parameter enters squared and its sign is a matter of convention.  Taking the positive root by comparing with the small-$r$ expansion of Eq.~\eqref{f_origin} gives
\begin{equation}
	H = \frac{1}{h}\sqrt{\frac{n+1}{n-2}} \; > 0.
	\label{H_n}
\end{equation}
The static OCK geometry determines only \(H^{2}\), not the sign of \(H\); the choice \(H>0\) is a convenient convention that selects the expanding de Sitter branch.  What matters for the thermodynamic discussion is that the inner horizon carries a positive geometric entropy \(S_{\rm inner}=\pi h_{c}^{2}\), irrespective of how the global time orientation is fixed. This is a crucial distinction from the black-hole gravastar scenario, where the de Sitter core is contracting ($H<0$) and carries negative entropy \cite{Volovik:2023phl}.

For $n>2$, the metric function $f(r)$ possesses a second zero at $r=h_{\rm c}<h$, corresponding to an inner (Cauchy) horizon. For the illustrative case $n=3$,
\begin{equation}
	h_{\rm c}(n=3) = \frac{1+\sqrt{13}}{6}\,h \approx 0.7676\,h.
	\label{n3_horizon}
\end{equation}

\subsection{Entropy release during collapse}

For any Killing horizon, including the inner horizon, the Wald--Noether-charge formalism gives the entropy~\cite{Wald:1993nt,Iyer:1994ys,Wald:1999vt}
\begin{equation}
	S_{\rm inner} = \frac{A_{\rm inner}}{4} = \pi h_{\rm c}^2.
	\label{S_inner}
\end{equation}

This entropy is \emph{hidden} from external observers in equilibrium: the inner horizon is causally disconnected from the outside world, so its area does not contribute to the entropy measured at infinity. However, when the black hole undergoes a transition that destroys the inner horizon, this entropy, assuming the validity of the generalized second law, must be released to the environment.

The total entropy of the regular black hole is therefore
\begin{equation}
	S_{\rm reg} = S_{\rm outer} + S_{\rm inner} = \frac{A}{4} + \pi h_{\rm c}^2,
	\label{Sreg}
\end{equation}
where $A=4\pi h^2$ is the area of the event horizon.

During gravitational collapse, the parameter $n$ decreases monotonically, $\dot n(v)<0$, driven by the null convergence condition~\cite{Ovalle:2026lxb}. As $n$ decreases from a regular value ($n>2$) toward zero, the inner horizon shrinks continuously and the entropy stored in it decreases correspondingly. The entropy difference between the regular black hole and the Schwarzschild state is
\begin{equation}
	\Delta S(n) = S_{\rm reg}(n) - S_{\rm Sch} = \pi h_{\rm c}^{2}.
	\label{DeltaS}
\end{equation}

For the illustrative case $n=3$,
\begin{equation}
	\Delta S(n=3) = \pi h_{\rm c}^2 \approx 0.589\,\pi h^2 = 0.589\,\frac{A}{4}.
	\label{deltaS_n3}
\end{equation}
Thus, a regular black hole with $n=3$ stores an additional entropy equal to approximately $59\%$ of the Bekenstein--Hawking entropy of the event horizon.

The inner horizon survives down to $n=0$, where Minkowski breaking occurs: the metric function at the origin jumps discontinuously to $f(0) = -1/2 \neq 1$, signalling a loss of local Minkowski structure. This classical obstruction indicates that the transition from a regular black hole to the Schwarzschild state must be fundamentally quantum in nature.

\subsection{Concept of gravitational collapse}
\label{subsec:OCK_collapse}

A potential source of confusion in the OCK framework is the apparent tension between the static nature of the geometry and the concept of gravitational collapse. The OCK metric is manifestly static, describing the equilibrium state of a regular black hole with an expanding de Sitter core. Yet the entropy release mechanism relies on a dynamical process in which the parameter $n$ decreases in time. Here we resolve this apparent paradox by carefully distinguishing between the static geometry and the dynamical evolution that drives the system from one static configuration to another.

The gravitational collapse is not encoded in the metric itself but in the time dependence of $n$. In the dynamical setting, one promotes $n$ to a function of the advanced time $v$, so that $n=n(v)$. The null convergence condition then imposes $\dot n(v)<0$, meaning that $n$ decreases monotonically during collapse \cite{Ovalle:2026lxb}. This evolution drives the system from one static configuration (with a given $n$) to another (with a smaller $n$). Classically the parameter can only reach $n=0$; the exact Schwarzschild metric, corresponding to the limit $n=-1$ of the OCK family, lies beyond the Minkowski breaking point and would require a quantum jump to be realised.

The expanding de Sitter core ($H>0$) is a feature of each static snapshot, not a contradiction with the collapse dynamics. The collapse is defined by the evolution of $n$, not by the sign of $H$. As $n$ decreases, the inner horizon radius $h_c(n)$ shrinks, and the entropy stored in it is gradually released.

One may speculate that a full quantum gravitational process at the Minkowski breaking point could flip the sign of the effective cosmological constant, converting the expanding interior into a contracting de Sitter phase---the configuration that characterizes a black‑hole gravastar in Volovik's framework.  However, within the purely classical OCK geometry the core ceases to be de Sitter for $n<2$ (the effective cosmological constant becomes negative) and no such transition is implied.  The two frameworks are therefore complementary paradigms rather than successive stages of a single classical collapse; their connection, if any, must be quantum in nature.

\subsection{Quantum breakdown and critical value of $n$}

The inner horizon surface gravity is defined as $\kappa_{\rm inner} = -\frac12 f'(h_{\rm c})$, which is positive for the Cauchy horizon.  Using the metric function~\eqref{f_N1} we obtain
\begin{equation}
	\kappa_{\rm inner} = \frac{1}{h(n-2)}\left[ (n+1)\frac{h_{\rm c}}{h} - \frac{3n}{2}\left(\frac{h_{\rm c}}{h}\right)^{n-1} \right].
	\label{kappa}
\end{equation}

As $n\to 0^{+}$, the inner horizon radius vanishes as $h_c/h \simeq \left(2/3\right)^{1/n}$ and the surface gravity diverges as
\begin{equation}
	\kappa_{\rm inner} \sim \frac{n}{2h}\left(\frac{3}{2}\right)^{1/n}.
	\label{kappa_divergence}
\end{equation}

The semiclassical description breaks down when the inner-horizon temperature exceeds the Planck temperature, $\kappa_{\rm inner} \sim 1/\ell_P$. Solving this condition yields $n \sim [\ln(h/\ell_P)]^{-1}$.
For astrophysical black holes, $h/\ell_P \sim 10^{38}$, giving
$n \sim  0.01$, which is deep in the quantum regime.\\


The OCK framework describes entropy release during gravitational collapse as a sequence of stages that naturally emerges from the geometry and its dynamics. The process begins with an initial regular black hole characterized by an expanding de Sitter core ($H>0$) carrying positive inner-horizon entropy \(S_{\rm inner}=\pi h_c^2\). As the parameter \(n\) decreases monotonically, the inner horizon shrinks continuously, and the stored entropy is gradually released. When the system reaches \(n\sim 0.01\), the semiclassical description breaks down due to the diverging inner-horizon temperature, signaling the onset of the quantum regime. The Minkowski breaking at \(n=0\) then marks a kinematic obstruction that requires a quantum resolution, possibly linking to the contracting de Sitter phase of a gravastar, but no continuous classical transition exists. The final state of the process is the Schwarzschild black hole with entropy \(A/4\), with the released entropy carried away by radiation to satisfy the generalized second law.

This picture will be complemented in the next section by a framework in which a contracting de Sitter core carries negative entropy that cancels the horizon entropy. The fundamental distinction between the two frameworks lies in the sign of the Hubble parameter: expanding cores yield positive entropy, while contracting cores yield negative entropy. Both frameworks, however, embody the same thermodynamic principle: the shrinking of the interior region drives the system toward increased total entropy.

\section{Local Thermodynamics of de Sitter Vacuum}
\label{sec:Volovik}

Volovik's local thermodynamics of the de Sitter vacuum provides a complementary perspective on gravitational entropy, based on the observation that the de Sitter spacetime acts as a thermal bath for matter with a well-defined local temperature. In this section, we review the key elements of this framework, emphasizing the sign of the entropy and its relation to the expansion or contraction of the de Sitter state.

\subsection{Local temperature and its physical origin}

Volovik established that the de Sitter vacuum serves as a heat bath for matter with the local activation temperature $T = H/\pi$~\cite{Volovik:2023phl}, which is twice the Gibbons-Hawking temperature $T_{\rm GH}=H/2\pi$~\cite{Gibbons:1977mu}. This temperature governs local processes that take place well inside the cosmological horizon and have no direct relation to the horizon itself.

The physical origin of this temperature lies in the specific symmetry of the de Sitter spacetime. The de Sitter metric in Painlev\'e-Gullstrand form is
\begin{equation}
	ds^2 = -dt^2 + (dr - Hr\,dt)^2 + r^2 d\Omega^2,
	\label{PG}
\end{equation}
which is invariant under the modified translations $\mathbf r \rightarrow \mathbf r - e^{Ht}\mathbf a$.
This symmetry, which reduces to ordinary translations in the limit $H\to 0$, ensures that all comoving observers at any point of the de Sitter space perceive the same local temperature. Thus, the de Sitter vacuum is unique among gravitational backgrounds in providing a thermal bath whose temperature does not violate the spacetime symmetry.

The temperature $T=H/\pi$ governs activation processes that are energetically forbidden in Minkowski spacetime but allowed in the de Sitter background. For example, the ionization rate of an atom in the de Sitter environment is $w \sim \exp\left(-E/T\right)$, where $E$ is the ionization potential. Similarly, the decay of a composite particle into two particles with total mass exceeding the original mass is governed by the same temperature~\cite{Volovik:2023phl}. Note that for $H<0$ the temperature becomes negative, which is consistent with the population inversion characteristic of negative-temperature systems; the absolute value $|T|$ still controls the activation rates.

\subsection{Local entropy density}

From the Friedmann equations of general relativity, the vacuum energy density is
\begin{equation}
	\epsilon_{\rm vac} = \frac{3}{8\pi}H^2 = \frac{3\pi}{8}T^2.
	\label{epsilon}
\end{equation}

Assuming that the temperature $T=H/\pi$ is the local temperature of the de Sitter vacuum, we can determine the free energy density. In the de Sitter state, the free energy density is quadratic in temperature:
\begin{equation}
	F = \epsilon_{\rm vac} - T s_{\rm vac}.
	\label{F_def}
\end{equation}
Using the thermodynamic relation $s_{\rm vac} = -\partial F/\partial T$, we obtain the local entropy density
\begin{equation}
	s_{\rm vac} = -\frac{\partial F}{\partial T} = \frac{3\pi}{4}T = \frac{3}{4}H.
	\label{s_local}
\end{equation}

Note that Eq.~\eqref{s_local} shows that the sign of the entropy density follows the sign of the Hubble parameter:
\begin{equation}
	s_{\rm vac} = \begin{cases}
		\text{positive}, & H > 0 \quad \text{(expanding de Sitter)},\\
		\text{negative}, & H < 0 \quad \text{(contracting de Sitter)}.
	\end{cases}
	\label{s_sign}
\end{equation}

This is a remarkable result: the entropy of the de Sitter vacuum is not an intrinsic positive quantity but changes sign with the direction of expansion. An expanding de Sitter universe ($H>0$) has positive entropy, while a contracting de Sitter universe ($H<0$) has negative entropy. This sign change is central to the thermodynamic interpretation of black-hole gravastars and the complementarity with the OCK framework.

\subsection{Modified Gibbs-Duhem relation}

The conventional vacuum pressure $P_{\rm vac}$ obeys the equation of state $w=-1$ and enters the energy-momentum tensor as
\begin{equation}
	T^{\mu\nu}= \Lambda g^{\mu\nu} = \mathrm{diag}(\epsilon_{\rm vac}, P_{\rm vac}, P_{\rm vac}, P_{\rm vac}),
	\label{EnergyMomentum}
\end{equation}
where $P_{\rm vac}=-\epsilon_{\rm vac}$. In the de Sitter state, $P_{\rm vac}<0$.

This pressure does not satisfy the standard thermodynamic Gibbs-Duhem relation $Ts_{\rm vac}=\epsilon_{\rm vac}+P_{\rm vac}$, because the right-hand side is zero. The reason is that this relation does not account for the gravitational degrees of freedom. Earlier work has shown that gravity contributes to thermodynamics with a pair of thermodynamically conjugate variables: the gravitational coupling $K=1/16\pi$ (in units $G=1$) and the scalar curvature $\mathcal R$~\cite{Klinkhamer:2008ff,Volovik:2021iim,Volovik:2020qtp}. The Einstein-Hilbert action contains the term $K\mathcal R$, and its contribution to thermodynamics is analogous to work density~\cite{Hayward:1997jp,Hayward:1998ee,Jacobson:1995ab,Nojiri:2023wzz}.

Including these gravitational variables, the modified Gibbs-Duhem relation takes the form
\begin{equation}
	T s_{\rm vac} = \epsilon_{\rm vac} + P_{\rm vac} - K\mathcal R,
	\label{GibbsDuhem}
\end{equation}
where $\mathcal R=-12H^2$ for the de Sitter spacetime. This equation is satisfied identically, since $\epsilon_{\rm vac}+P_{\rm vac}=0$ and $-K\mathcal R=12KH^2=Ts_{\rm vac}$.

Equation \eqref{GibbsDuhem} suggests the introduction of an effective pressure that absorbs the gravitational degrees of freedom:
\begin{equation}
	P = P_{\rm vac} - K\mathcal R.
	\label{EffectiveP}
\end{equation}
Then the conventional Gibbs-Duhem relation is restored:
\begin{equation}
	Ts_{\rm vac} = \epsilon_{\rm vac} + P.
	\label{EffectiveGibbs}
\end{equation}

The effective de Sitter pressure $P$ is positive, $P=\epsilon_{\rm vac}>0$, and satisfies the equation of state $w=1$, which is characteristic of stiff matter introduced by Zel'dovich~\cite{Zeldovich:1961sbr}. Thus, due to the gravitational degrees of freedom, the de Sitter state behaves thermodynamically like a non-relativistic Fermi liquid, where the thermal energy is proportional to $T^2$, and like relativistic stiff matter with sound speed equal to the speed of light. This effective pressure reveals the thermodynamic nature of the de Sitter vacuum as stiff matter ($w=1$), a key property that will be used in the unified framework developed in Sec.~\ref{sec:unification}. In that framework, the sign of the entropy, determined by the sign of the Hubble parameter, provides the direct link between the expanding core of the OCK black hole and the contracting core of the black-hole gravastar.

\subsection{Total entropy in the Hubble volume}

The cosmological horizon of a de Sitter spacetime is located at the radius $R_H = 1/|H|$, enclosing the volume $V_H = \frac{4\pi}{3} R_H^3 = \frac{4\pi}{3|H|^3}$. Integrating the local entropy density~\eqref{s_local} over this volume gives the total gravitational entropy of the de Sitter vacuum:
\begin{equation}
	S_{\rm dS} = s_{\rm vac} V_H = \frac{3}{4}H \cdot \frac{4\pi}{3|H|^3}
	= \frac{\pi H}{|H|^3} = \operatorname{sgn}(H)\,\frac{\pi}{H^2}.
	\label{S_dS}
\end{equation}
Using the horizon area $A = 4\pi R_H^2 = 4\pi/H^2$, this can be written in the more familiar form
\begin{equation}
	S_{\rm dS} = \operatorname{sgn}(H)\,\frac{A}{4}.
	\label{S_dS_A}
\end{equation}
Thus, for an expanding de Sitter phase ($H>0$) the total entropy is the Gibbons--Hawking entropy $A/4$, whereas for a contracting phase ($H<0$) it is \emph{negative}: $S_{\rm dS} = -A/4$.

This result reproduces the holographic bulk-surface correspondence: the entropy obtained from the local vacuum thermodynamics inside the horizon exactly matches the entropy attributed to the horizon itself, up to the sign given by the direction of expansion. Equation~\eqref{S_dS} (or its dimensional form $S_{\rm dS}= \operatorname{sgn}(H)\pi/(G H^2)$ when $G$ is restored) is the correct unified entropy function for a homogeneous de Sitter region; earlier versions of this work contained an erroneous expression $S(H)\propto -1/(4H)$, which is dimensionally inconsistent and gives the wrong sign. This sign-dependent entropy provides the foundation for the unified entropy dynamics, where the sign of the entropy follows the sign of the Hubble parameter.

\subsection{Gravastars and entropy cancellation}

In Volovik's framework, the term ``gravastar'' refers to a black hole that possesses a de Sitter core inside its event horizon, which distinguishes it from the horizonless gravastars originally proposed in Refs.~\cite{Chapline:2000en,Mazur:2001fv,Mottola:2023jxl}. In such a black-hole gravastar, the de Sitter core is contracting, as follows from the continuity of the shift vector in Painlev\'e-Gullstrand coordinates across the horizon. Consequently, with \(H<0\), the local entropy density is negative:
\begin{equation}
	s_{\rm core} = \frac{3}{4}H < 0.
	\label{s_core}
\end{equation}

The total entropy of the de Sitter region inside the horizon is obtained from~\eqref{S_dS_A} with $H<0$:
\begin{equation}
	S_{\rm core} = -\frac{A}{4}.
	\label{S_core}
\end{equation}

This negative entropy cancels the positive Bekenstein-Hawking entropy of the black hole horizon, yielding
\begin{equation}
	S_{\rm gravastar} = S_{\rm BH} + S_{\rm core} = \frac{A}{4} - \frac{A}{4} = 0.
	\label{S_gravastar}
\end{equation}

Thus, the black-hole gravastar has zero total entropy and, consistently, no Hawking radiation. This entropy cancellation encapsulates the key distinction between the two frameworks: a contracting de Sitter core carries negative entropy that cancels the horizon entropy, while an expanding core carries positive entropy that is released during collapse. In both cases, the sign of the Hubble parameter determines the thermodynamic role of the core. This complementary relationship will be formalized in Sec.~\ref{sec:unification}, where we show how the sign of $H$ provides the direct link between the expanding core of the OCK black hole and the contracting core of the black-hole gravastar.

These results establish the thermodynamic foundation for the complementarity with the OCK framework. The key ingredients are the local temperature \(T=H/\pi\) (which can be negative), the entropy density \(s_{\rm vac}=3H/4\) whose sign follows the sign of \(H\), the sign-dependent total entropy \(S_{\rm dS}= \operatorname{sgn}(H)\,A/4\), and the cancellation mechanism for black-hole gravastars. Together, these elements provide the essential thermodynamic structure that we will exploit in the next section to develop the unified description of gravitational collapse.

\section{Complementarity and Unification}
\label{sec:unification}

Having reviewed the OCK framework of entropy release from regular
Schwarzschild black holes and Volovik's local thermodynamics of the de
Sitter vacuum, we now establish the complementarity between these two
approaches. The key observation is that both frameworks describe a
closely related thermodynamic principle: the reduction of the interior
contribution is associated with an increase of the total entropy, but
from opposite perspectives---in the OCK framework through the
disappearance of a positive inner-horizon entropy, and in Volovik's
description through the cancellation of a negative bulk entropy. The thermodynamic correspondence between these descriptions is naturally organized by the sign of the Hubble parameter associated with the de Sitter sector.

The geometrical origin of this sign distinction becomes particularly transparent by comparing the static and Painlev\'e--Gullstrand representations of de Sitter spacetime. In static coordinates,
	\begin{equation}
		ds^2=-(1-H^2r^2)dt^2
		+\frac{dr^2}{1-H^2r^2}+r^2d\Omega^2 ,
		\label{eq:dS_static_unification}
	\end{equation}
the geometry depends only on $H^2$ and therefore does not distinguish between expanding and contracting de Sitter patches. By contrast, in Painlev\'e--Gullstrand coordinates,
	\begin{equation}
		ds^2=-dT^2+\left(dr-v(r)dT\right)^2+r^2d\Omega^2,
		\qquad v(r)=\pm Hr ,
		\label{eq:dS_PG_unification}
	\end{equation}
the two signs explicitly encode opposite time orientations: $v=+Hr$ describes the expanding patch, whereas $v=-Hr$ describes the contracting one.
	
This distinction has a direct thermodynamic counterpart. The horizon area depends only on $H^2$ and is therefore the same for the expanding and contracting descriptions,
	\begin{equation}
		A_H=\frac{4\pi}{H^2},
	\end{equation}
	whereas the sign-sensitive de Sitter entropy may be written as
	\begin{equation}
		S_{\rm dS}
		=\operatorname{sgn}(H)\frac{A_H}{4}
		=\operatorname{sgn}(H)\frac{\pi}{H^2}.
		\label{eq:dS_oriented_entropy}
	\end{equation}
Geometrically, this sign may be interpreted in terms of the orientation
of the horizon generator: reversing the time orientation reverses the
horizon-generating Killing field, $\xi^a\rightarrow-\xi^a$, and hence
the associated oriented Noether charge. In this interpretation, the
expanding and contracting de Sitter sectors possess the same positive
horizon area but opposite oriented entropy. The entropy sign therefore
distinguishes expansion from contraction rather than a change in the
intrinsic de Sitter geometry itself. This provides a geometrical basis
for the sign-sensitive thermodynamic distinction underlying the
OCK--Volovik comparison developed below.

\subsection{Expanding versus contracting de Sitter}

\begin{table*}[ht!]
	\centering
	\caption{Comparison of the OCK and Volovik frameworks}
	\label{tab:comparison}
	\begin{tabular}{p{0.23\textwidth}|p{0.35\textwidth}|p{0.35\textwidth}}
		\toprule
		\textbf{Aspect} & \textbf{OCK Framework} & \textbf{Volovik Framework} \\
		\midrule
		Core type & Expanding de Sitter ($n>2$) & Contracting de Sitter \\
		Hubble parameter & $H>0$ (for $n>2$) & $H<0$ \\
		Extra entropy & Positive: $+\pi h_c^2$ (inner horizon) &
		Negative: $-A/4$ (bulk) \\
		Physical process & Entropy release as inner horizon shrinks &
		Entropy cancellation (horizon $+$ core $=0$) \\
		Regime of validity & Classical evolution down to $n=0$; core
		becomes AdS for $n<2$ & Static configuration (by construction) \\
		Quantum transition & Minkowski breaking at $n=0$
		(kinematic obstruction) & Not emphasized \\
		Geometric setting & Regular Schwarzschild interior &
		Black-hole gravastar \\
		\bottomrule
	\end{tabular}
\end{table*}

In the OCK geometry, the $n>2$ core is locally de Sitter and may be
associated with the expanding branch, $H>0$, whereas Volovik's
black-hole gravastar employs the contracting branch, $H<0$. The
thermodynamic significance of this distinction follows from Volovik's
local relation
\begin{equation}
	s_{\rm vac}=\frac{3H}{4},
\end{equation}
which assigns positive bulk entropy density to expanding de Sitter and
negative bulk entropy density to contracting de Sitter.

The entropy quantities entering the two constructions should,
nevertheless, be distinguished carefully. In the OCK geometry the
additional positive contribution is the geometric entropy
$S_{\rm inner}=\pi h_c^2$ associated with the inner Killing horizon,
whereas in Volovik's construction the negative contribution is a bulk
vacuum entropy obtained by integrating $s_{\rm vac}$ over the
contracting de Sitter core. They are therefore not identical entropy
functionals. Their complementarity lies instead in the thermodynamic
role they play: reducing the interior contribution increases the total
entropy, through the release of a positive contribution in the OCK
description and through the removal of a negative contribution in
Volovik's description.

\subsection{Entropy dynamics in the two paradigms}

Despite their apparent differences, both frameworks embody the same thermodynamic principle: the shrinking of the interior region drives the system toward increased total entropy. This manifests differently in each scenario: in the OCK framework, as the inner horizon shrinks, positive entropy is released; in Volovik's framework, a contracting de Sitter core carries negative entropy that cancels the positive horizon entropy.

Consider the entropy changes in each scenario. In the OCK case, the regular black hole has entropy $S_{\rm reg}=A/4+\pi h_c^2$, while the Schwarzschild state has $S_{\rm Sch}=A/4$. The black hole entropy alone decreases by $\Delta S_{\rm BH}=-\pi h_c^2<0$, but the collapse is non-adiabatic: the released entropy is carried away by radiation, $\Delta S_{\rm rad}\ge\pi h_c^2$. The total entropy change therefore satisfies $\Delta S_{\rm total}=-\pi h_c^2+\Delta S_{\rm rad}\ge0$, preserving the generalized second law. In the Volovik case, the gravastar has zero total entropy, while the black hole state has entropy $S_{\rm BH}=A/4$. The entropy change is $\Delta S_{\rm total}=A/4 > 0$, with the entropy increasing directly as the negative-entropy core shrinks.

The two scenarios are thus complementary: in OCK, the black hole entropy decreases while the environment's entropy increases; in Volovik, the total entropy increases as the negative-entropy core disappears. Both are consistent with the generalized second law. This complementarity is traced to the sign of the Hubble parameter: the expanding core releases positive entropy, while the contracting core has negative entropy that cancels the horizon entropy.

\begin{table*}[ht!]
	\centering
	\caption{Comparison of entropy dynamics in the two frameworks}
	\label{tab:entropy_dynamics}
	\begin{tabular}{p{0.28\textwidth}|p{0.32\textwidth}|p{0.32\textwidth}}
		\toprule
		\textbf{Aspect} & \textbf{OCK Scenario} & \textbf{Volovik Scenario} \\
		\midrule
		Initial state & Regular BH: $S_{\rm reg}=A/4+\pi h_c^2$ & Gravastar: $S_{\rm gravastar}=0$ \\
		Final state & Schwarzschild: $S_{\rm Sch}=A/4$ & Black hole: $S_{\rm BH}=A/4$ \\
		Entropy change (system) & $\Delta S_{\rm BH}=-\pi h_c^2<0$ & $\Delta S_{\rm total}=A/4>0$ \\
		Entropy conservation & $\Delta S_{\rm BH}+\Delta S_{\rm rad}\ge0$ & Direct increase \\
		Mechanism & Positive entropy release & Negative entropy cancellation \\
		\bottomrule
	\end{tabular}
\end{table*}

\subsection{Minkowski breaking and the limits of classical unification}

It is tempting to regard the Minkowski breaking at $n=0$ in the OCK analysis as a continuous transition from an expanding de Sitter core to a contracting one, thereby unifying the two frameworks into a single collapse sequence.  However, this interpretation is not supported by the classical OCK geometry.

For $n>2$ the interior is a genuine expanding de Sitter core, and the inner horizon carries positive entropy.  As $n$ decreases, the effective cosmological constant $\Lambda_{\rm eff}(n)=3(n+1)/[h^{2}(n-2)]$ increases, diverging as $n\to2^{+}$.  For $n<2$, $\Lambda_{\rm eff}$ becomes negative, and the core is anti-de Sitter---not de Sitter.  Therefore, the local de Sitter thermodynamics of Volovik (which relies on a positive cosmological constant and a well-defined Hubble parameter) does not apply in the regime $n<2$.  The inner horizon continues to shrink and its geometric entropy is still released, but the process is no longer described by a de Sitter core with $H>0$.

At $n=0$, the metric function develops a discontinuity (Minkowski breaking), signalling that the Schwarzschild point mass cannot be formed classically.  This is a kinematic obstruction, not a mechanism that converts the interior into a contracting de Sitter phase.  Consequently, the OCK regular black hole and the Volovik black-hole gravastar are complementary paradigms that share the same sign‑dependent entropy principle, but they are \emph{not} successive stages of a single classical collapse.

\subsection{Speculative quantum connection}

Although the classical OCK solution does not evolve into a contracting de Sitter gravastar, one may speculate that a full quantum gravitational process at the Minkowski breaking point could effectively flip the sign of the effective cosmological constant, converting the expanding interior into a contracting de Sitter phase.  The divergence of the inner-horizon temperature as $n\to0^{+}$ and the breakdown of the semiclassical description at $n\sim 1/\ln(h/\ell_{P})$ indicate that quantum effects dominate near the breaking point.  If an instanton exists that interpolates between an expanding de Sitter interior and a contracting one, the Minkowski breaking would indeed correspond to a quantum phase transition, and the gravastar would emerge as an intermediate state with zero total entropy before the core eventually collapses to the Schwarzschild singularity.  In such a scenario, the total entropy would flow as
\[
S = \frac{A}{4} + \pi h_{c}^{2} \;\longrightarrow\; 0 \;\longrightarrow\; \frac{A}{4},
\]
Such a sequence can satisfy the generalized second law provided the first step is accompanied by radiation (or other environmental degrees of freedom) that carries an entropy at least \(\frac{A}{4}+\pi h_{c}^{2}\), just as in the pure OCK scenario.  Without specifying this environmental contribution the sequence alone does not manifestly obey the second law. At present this remains a conjecture, and constructing the relevant instanton is an open problem.

\begin{figure*}[ht!]
	\centering
	\includegraphics[scale=0.47]{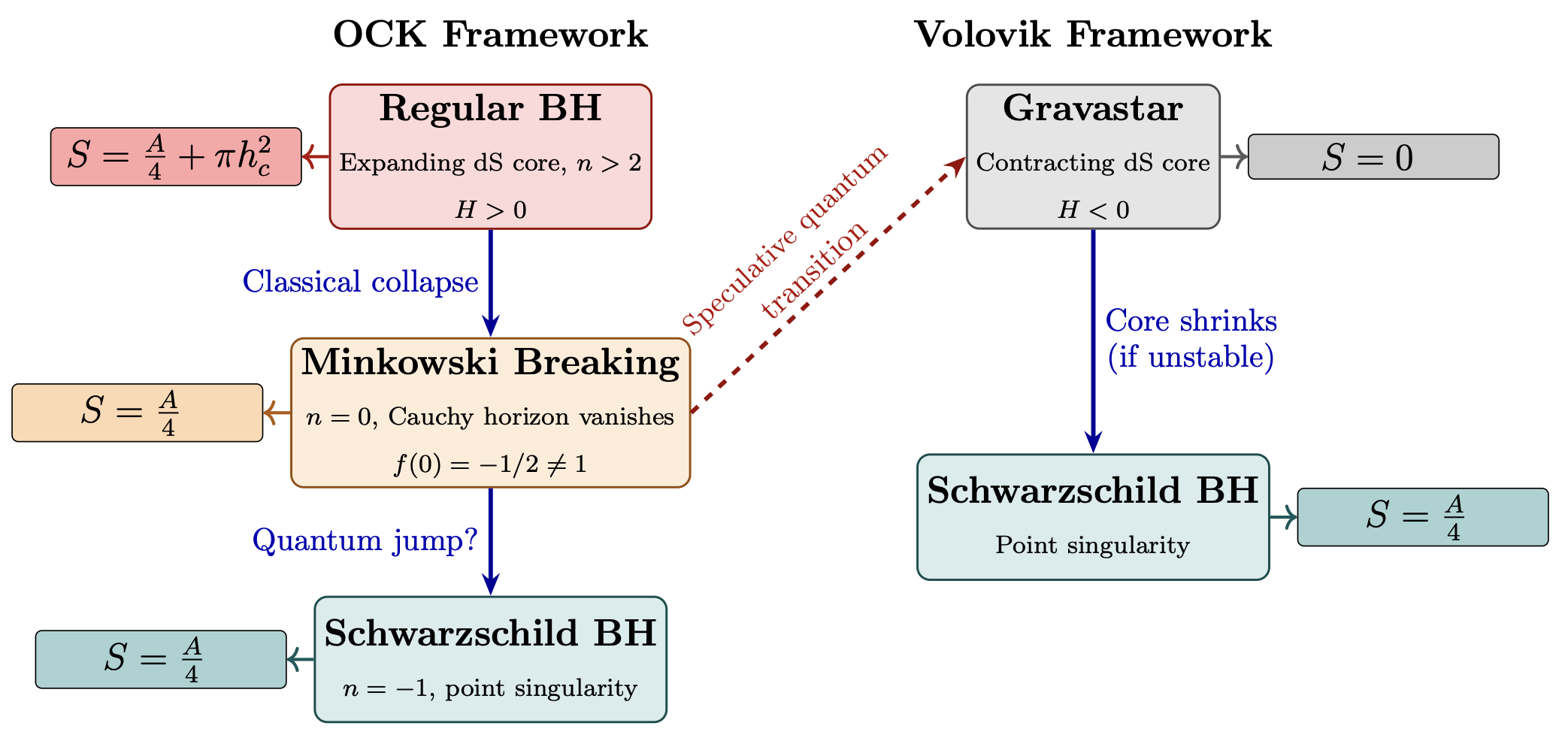}
	\caption{\textbf{Entropy flow in the two complementary frameworks.}
		\emph{Left column:} The OCK regular black hole contains an expanding de~Sitter core ($H>0$) and carries positive inner‑horizon entropy $S = A/4 + \pi h_c^2$.  Classical collapse drives the parameter $n$ to zero; at $n=0$ the inner horizon disappears and the metric function undergoes a Minkowski breaking, leaving $S = A/4$.  Whether a further (quantum) jump to the Schwarzschild state is required remains an open question.
		\emph{Right column:} In Volovik's picture, a black‑hole gravastar possesses a contracting de~Sitter core ($H<0$) whose negative bulk entropy exactly cancels the horizon entropy, giving $S = 0$.  If the core shrinks (e.g. due to perturbations), the entropy rises to $S = A/4$, restoring the Schwarzschild black hole.
		The dashed red arrow indicates a \emph{speculative} quantum transition that could connect the two paradigms by flipping the sign of the Hubble parameter at the Minkowski breaking point.  No concrete dynamical mechanism for such a transition is known.  Both branches satisfy the generalized second law.}
	\label{fig:entropy_flow}
\end{figure*}

\subsection{Summary}

The two frameworks are complementary paradigms governed by the same sign-dependent entropy principle (see Fig.~\ref{fig:entropy_flow}): an expanding de Sitter core ($H>0$) carries positive extra entropy that is released during collapse (the OCK scenario), whereas a contracting de Sitter core ($H<0$) carries negative entropy that exactly cancels the horizon entropy, yielding a zero-entropy gravastar (Volovik's scenario). Both satisfy the generalized second law. The Minkowski breaking at $n=0$ is a classical kinematic obstruction, not a continuous transition to a contracting phase; any dynamical connection between the two pictures would require a quantum mechanism that has yet to be identified. The sign of the Hubble parameter thus acts as the unifying order parameter that dictates the thermodynamic role of the interior.

\section{Thermodynamic analysis of the entropy flow}
\label{sec:analysis}

The complementarity between the OCK and Volovik frameworks can be sharpened by two concrete analyses: (i) tracking the entropy released as the inner horizon shrinks, both for classical continuous evolution and for a discrete quantum cascade, and (ii) formulating a Landau-type thermodynamic potential that exposes the role of the Hubble parameter as an order parameter distinguishing the expanding and contracting de Sitter phases.

\subsection{Entropy-release profile in the OCK collapse}
\label{subsec:entropy_profile}

For the $N=1$ OCK solution, the inner-horizon entropy is $S_{\rm inner}(n)=\pi h_{c}^{2}(n)$, where $h_{c}(n)$ is the smaller positive root of $f(r)=0$ in Eq.~\eqref{f_N1}.  Although a closed analytic form for $h_{c}$ is available only for special values of $n$, the function is easily computed numerically.  Table~\ref{tab:entropy_profile} lists $h_{c}/h$ and the entropy fraction $\Delta S/(A/4) = (h_{c}/h)^{2}$ for several values of the (static) parameter $n$.

\begin{table}[h]
	\caption{Inner-horizon radius and entropy fraction for the $N=1$ OCK solution.}
	\label{tab:entropy_profile}
	\begin{ruledtabular}
		\begin{tabular}{c c c}
			$n$ & $h_{c}/h$ & $\Delta S/(A/4)$ \\
			\hline
			$3$   & $0.7676$ & $0.589$ \\
			$4$   & $0.8165$ & $0.667$ \\
			$5$   & $0.8484$ & $0.720$ \\
			$6$   & $0.871$  & $0.759$ \\
			$7$   & $0.8875$ & $0.788$ \\
			$10$  & $0.919$  & $0.845$ \\
			$100$ & $0.992$  & $0.984$ \\
			$\infty$ & $1$      & $1$ \\
		\end{tabular}
	\end{ruledtabular}
\end{table}

During a classical continuous collapse, $n(v)$ evolves monotonically from a large initial value toward zero.  The inner horizon shrinks and $S_{\rm inner}$ decreases smoothly; the released entropy is carried away by radiation, so the generalised second law is satisfied at every instant.  The entropy release per unit decrease of $n$ is $dS_{\rm inner}/dn$, which remains finite for all $n>0$.

If, instead, the transition between the static integer-$n$ equilibrium states occurs via quantum jumps, the entropy is emitted in finite steps as the parameter decreases from one integer to the next.  Let us define the step from $n+1$ to $n$ (with $n=3,4,5,\dots$) as
\begin{equation}
	\Delta S_{n} = S_{\rm inner}(n+1)-S_{\rm inner}(n) \qquad (n\ge 3).
	\label{deltaS_step}
\end{equation}

Using the values in Table~\ref{tab:entropy_profile} one finds
\[
\begin{aligned}
	\Delta S_{3} &\approx 0.078\,\frac{A}{4}, \qquad
	\Delta S_{4} \approx 0.053\,\frac{A}{4}, \\
	\Delta S_{5} &\approx 0.039\,\frac{A}{4}, \qquad
	\Delta S_{6} \approx 0.029\,\frac{A}{4},
\end{aligned}
\]
and the steps continue to diminish as \(n\) increases.
Each step becomes smaller as $n$ increases, reflecting the approach of the inner horizon to the event horizon.  A quantum cascade of this kind would produce a train of discrete pulses whose amplitudes are determined by the integer jumps, providing a possible observational signature of the discrete internal structure.  The sum of all integer steps from $n=\infty$ down to $n=3$ does not account for the full inner-horizon entropy; a final, larger jump from $n=3$ to $n=0$ (where the inner horizon disappears) would release the remaining fraction $S_{\rm inner}(3) \approx 0.589\,A/4$.  In the continuous evolution picture the integral of $dS_{\rm inner}/dn$ from $n=0$ to $n=\infty$ recovers the full inner-horizon entropy $\pi h^{2}=A/4$.

\subsection{Landau-type potential and order parameter}
\label{subsec:Landau}

The sign of the Hubble parameter controls the sign of the total de Sitter entropy, $S_{\rm dS}= \operatorname{sgn}(H)\,A/4$.  This suggests that $H$ can be viewed as an order parameter that distinguishes two different thermodynamic phases.  To make this analogy concrete, we construct a phenomenological Landau-type free energy $F(H)$ whose minimum determines the equilibrium state and whose barrier structure encodes the (speculative) quantum transition between the phases.

For a homogeneous de Sitter region of Hubble radius $R_{H}=1/|H|$, the internal energy and entropy are
\[
E = \epsilon_{\rm vac} V_{H} = \frac{3}{8\pi}H^{2}\cdot\frac{4\pi}{3|H|^{3}} = \frac{1}{2|H|}, \qquad
S = \operatorname{sgn}(H)\,\frac{\pi}{H^{2}} .
\]
Using the local temperature $T=H/\pi$ (which can be positive or negative), the free energy $F = E - T S$ becomes
\begin{equation}
	F(H) = \frac{1}{2|H|} - \frac{H}{\pi}\cdot\operatorname{sgn}(H)\,\frac{\pi}{H^{2}}
	= \frac{1}{2|H|} - \frac{1}{|H|} = -\frac{1}{2|H|} .
	\label{F_H}
\end{equation}
Equation~\eqref{F_H} depends only on $|H|$ and is always negative, decreasing without bound as $|H|\to0$.  It does not exhibit a minimum at any finite $H$, which reflects the fact that a free energy constructed from the static de Sitter patch does not capture the dynamical stability of the de Sitter state.  To obtain a potential that distinguishes the sign of $H$, one must instead consider the entropy difference between configurations as the driver of quantum transitions.  A more useful quantity is therefore the \emph{entropic potential}
\begin{equation}
	V(H) \equiv -S(H) = -\operatorname{sgn}(H)\,\frac{\pi}{H^{2}} .
	\label{V_H}
\end{equation}
The probability of a quantum transition between two states with Hubble parameters $H_{i}$ and $H_{f}$ is expected to scale as $\Gamma \propto \exp\bigl[-(V(H_{f})-V(H_{i}))\bigr] = \exp\bigl[S(H_{f})-S(H_{i})\bigr]$, in agreement with the entropy enhancement discussed in the main text.

\begin{figure}[ht!]
	\centering
	\includegraphics[width=\columnwidth]{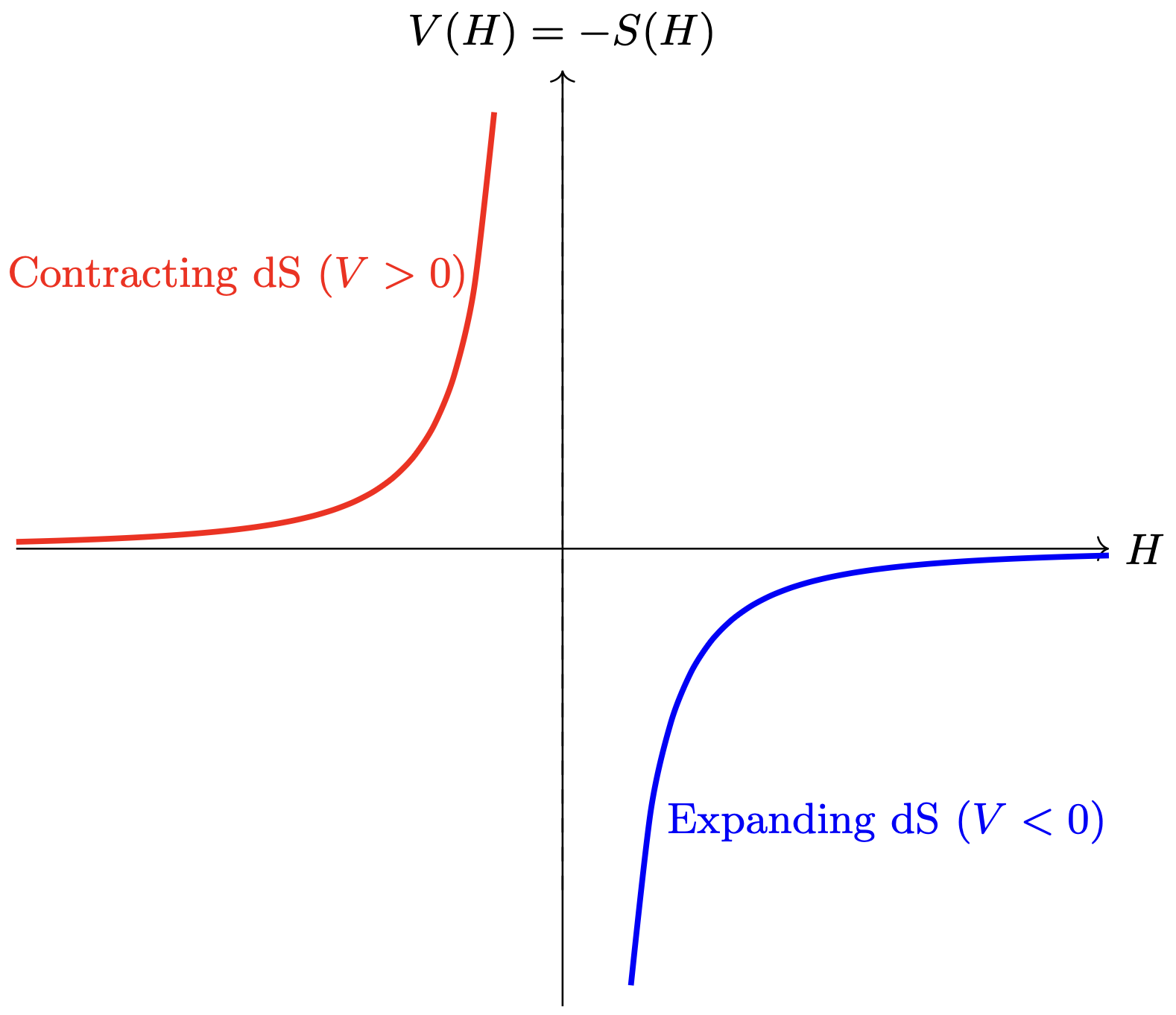}
	\caption{Entropic potential $V(H) = -S(H)$ as a function of the Hubble parameter. The expanding phase ($H>0$) has $V<0$, the contracting phase ($H<0$) has $V>0$. The divergence at $H=0$ separates the two phases by an infinite barrier. A transition between them would require quantum tunnelling.}
	\label{fig:Landau}
\end{figure}

The potential $V(H)$ is plotted schematically in Fig.~\ref{fig:Landau}.  For $H>0$, $V(H) = -\pi/H^{2}$ is negative and dives to $-\infty$ as $H\to0^{+}$; for $H<0$, $V(H) = +\pi/H^{2}$ is positive and rises to $+\infty$ as $H\to0^{-}$.  The two branches are separated by an infinite barrier at $H=0$, which cleanly separates the expanding and contracting de Sitter phases.  In this picture, the classical OCK collapse drives the system along the positive-$H$ branch toward larger $H$ (the inner horizon shrinks and the core entropy decreases), eventually leaving the regime where the de Sitter description applies.  A transition to the contracting de Sitter phase would require a quantum tunnelling event that traverses the classically forbidden region near $H=0$.  The necessity of such a tunnelling is consistent with the breakdown of the semiclassical approximation at $n\sim 1/\ln(h/\ell_{P})$ and with the discontinuity in the metric function at $n=0$.  While the shape of $V(H)$ is purely phenomenological, it illustrates how the sign-dependent entropy creates a barrier between the expanding and contracting phases, and it reinforces the notion that a quantum process is required to connect the two paradigms. The divergence at \(H=0\) reflects the infinite Hubble volume and should not be interpreted as a literal infinite tunnelling barrier.  The potential simply illustrates that the sign‑dependent entropy separates the expanding and contracting de Sitter phases; it does not by itself constitute a Landau free energy nor does it identify \(n=0\) with \(H=0\).

In summary, the explicit entropy-release profile quantifies the graduality (or discreteness) of the collapse, while the Landau-type entropic potential offers an intuitive thermodynamic picture of the order-parameter character of $H$ and the barrier that separates the expanding and contracting de Sitter phases.  Together they reinforce the main message of this work: the sign of the Hubble parameter governs the thermodynamics of gravitational collapse, and a complete dynamical unification of the expanding and contracting phases remains an open quantum problem.

\section{Unified Entropy Dynamics and Implications}
\label{sec:dynamics}

Having established the complementarity of the OCK and Volovik frameworks above, we now consolidate their insights into a unified thermodynamic picture. We stress that the two frameworks are complementary paradigms, not successive stages of a single classical collapse.  Any dynamical connection between them requires a quantum mechanism that is not yet understood. The central quantity in this complementarity is the Hubble parameter \(H\), which determines both the sign of the entropy and the direction of the thermodynamic flow.

\subsection{Sign‑dependent entropy and its consequences}

For a homogeneous de Sitter region, the total gravitational entropy is
\begin{equation}
	S_{\rm dS}(H) = \operatorname{sgn}(H)\,\frac{\pi}{H^{2}}
	= \operatorname{sgn}(H)\,\frac{A}{4},
	\label{S_dS_correct}
\end{equation}
where $A = 4\pi/H^{2}$ is the horizon area.  Expanding de Sitter ($H>0$) therefore has positive entropy $A/4$, while contracting de Sitter ($H<0$) has negative entropy $-A/4$.  This sign dependence is the fundamental insight that links the two frameworks.

In the OCK solution (for $n>2$), the interior is an expanding de Sitter core with $H>0$.  The associated positive extra entropy is carried by the inner Killing horizon:
\begin{equation}
	S_{\rm inner}(n) = \pi h_{c}^{2}(n) \qquad (n>0).
	\label{S_inner_OCK}
\end{equation}
As \(n\) decreases during collapse, the inner horizon shrinks and the positive entropy $\pi h_{c}^{2}$ is gradually released.  Because the OCK geometry is only approximately de Sitter, the exact relation between $S_{\rm inner}$ and $H$ is not a simple algebraic function of $H$; only in the limit $n\gg1$ (where $h_{c}\to h$) does $S_{\rm inner}$ approach the full de Sitter value $\pi/H^{2}=A/4$. For finite \(n\) the Wald inner‑horizon entropy and the homogeneous de Sitter bulk entropy are different (e.g., at \(n=3\) they differ by a factor \(\sim 2.36\)).  The asymptotic agreement at \(n\to\infty\) provides a conceptual bridge, but the two constructions probe different aspects of the geometry.  Establishing a common microscopic origin remains an open problem.

In Volovik's black‑hole gravastar, the interior is a contracting de Sitter core with $H<0$.  Its total entropy is exactly the negative value
\begin{equation}
	S_{\rm core} = -\frac{A}{4},
	\label{S_core_Volovik}
\end{equation}
which cancels the Bekenstein--Hawking entropy of the event horizon, yielding a zero‑entropy object.

Both mechanisms, namely, the release of positive extra entropy and the cancellation by negative core entropy, are consistent with the generalized second law and reflect the same principle: the shrinking of the interior region drives the system toward increased total entropy. The sign of the Hubble parameter simply dictates whether the extra entropy is positive (OCK) or negative (Volovik).

\subsection{Limits of classical unification and quantum connection}

The OCK solution does not provide a classical path to a contracting de Sitter gravastar.  For \(n>2\) the core is de Sitter, but for \(n<2\) the effective cosmological constant becomes negative and the core is anti-de Sitter; Volovik's local de Sitter thermodynamics does not apply.  The Minkowski breaking at \(n=0\) is a kinematic obstruction---a discontinuity in the metric function that signals the impossibility of forming a Schwarzschild point mass classically---not a mechanism that flips the sign of \(H\).  Therefore, the two frameworks are complementary paradigms, not successive stages of a single classical collapse.

Although no classical evolution connects the OCK regular black hole to the Volovik gravastar, a quantum transition at the Minkowski breaking point could, in principle, change the sign of the effective cosmological constant.  The divergence of the inner-horizon temperature and the breakdown of the semiclassical description at \(n\sim 1/\ln(h/\ell_{P})\) indicate that quantum effects dominate near \(n=0\).  If an instanton exists that interpolates between an expanding de Sitter interior and a contracting one, the Minkowski breaking would correspond to a quantum phase transition, and the black-hole gravastar would emerge as an intermediate state with zero total entropy.  The complete entropy flow would then be
\[
S = \frac{A}{4} + \pi h_{c}^{2} \;\longrightarrow\; 0 \;\longrightarrow\; \frac{A}{4},
\]
consistent with the generalized second law at every step, provided the environmental entropy is included as discussed in Sec.~\ref{sec:unification}.  We stress that this is a conjecture; constructing the relevant instanton remains an open problem.  Nevertheless, the existence of a barrier between the two phases, as illustrated by the entropic potential of Sec.~\ref{sec:analysis}, makes the idea of a quantum tunnelling event a natural and physically motivated possibility.

Even in the absence of such a quantum connection, both frameworks individually satisfy the generalized second law.  In the OCK scenario, the black hole entropy decreases during collapse, but the released entropy is carried away by radiation:
\begin{equation}
	\Delta S_{\rm BH} + \Delta S_{\rm env} = -\pi h_{c}^{2} + \Delta S_{\rm rad} \ge 0,
	\label{GSL_OCK}
\end{equation}
where \(\Delta S_{\rm rad} \ge \pi h_{c}^{2}\).  In Volovik's scenario, the total entropy increases directly as the negative-entropy core shrinks:
\begin{equation}
	\Delta S_{\rm total} = \frac{A}{4} - 0 = \frac{A}{4} > 0.
	\label{GSL_Volovik}
\end{equation}
Thus, whether the two paradigms are connected or not, the total entropy of the universe never decreases.

\subsection{Quantum breakdown and entropy quantization}

The Minkowski breaking at \(n=0\) is intrinsically quantum. The inner-horizon surface gravity diverges as \(n\to0^{+}\) (see Eq.~\eqref{kappa_divergence}), implying
\begin{equation}
	T_{\rm inner} = \frac{\hbar\kappa_{\rm inner}}{2\pi} \to \infty .
	\label{T_divergence}
\end{equation}
The semiclassical description breaks down at \(n \sim 1/\ln(h/\ell_{P})\), which for astrophysical black holes gives \(n\sim0.01\), deep in the quantum regime. The discontinuity in the metric function at \(n=0\) signals a loss of local Minkowski structure. This combination of divergences and discontinuities indicates that whatever resolves the Minkowski breaking must be a quantum gravitational process; a smooth classical transition is ruled out.

At the same time, the integer nature of the OCK parameter \(n\) implies a quantized entropy spectrum for the regular black holes,
\begin{equation}
	S_{\rm reg}(n) = \frac{A}{4} + \pi h_{c}^{2}(n) ,
	\label{S_quantized}
\end{equation}
with \(h_{c}(n)\) taking discrete values. This supports Bekenstein's conjecture that the horizon area is quantized~\cite{Bekenstein:1974jk}. In the present framework, the quantization arises from the discrete internal structure of the regular black hole. Each value of \(n\) corresponds to a different interior configuration, all of which share the same exterior Schwarzschild metric. The number of distinct regular states for a fixed macroscopic mass is given by the number of allowed integer tuples,
\begin{equation}
	\Omega(\mathcal M) = \#\{(n_1,\dots,n_N): 2<n_1<\dots<n_N\le n_{\rm max}\},
	\label{Omega}
\end{equation}
where \(n_{\rm max}\) is a cutoff imposed by quantum gravity. In the semiclassical limit, one therefore expects
\begin{equation}
	S_{\rm BH} = \ln\Omega(\mathcal M) .
	\label{S_BH_lnOmega}
\end{equation}
This offers a concrete statistical-mechanics proposal for the origin of black hole entropy, with the quantum breakdown near the Minkowski breaking point providing the natural cutoff that renders the microstate count finite.

\subsection{Summary and thermodynamic arrow of time}

The sign of the Hubble parameter determines the sign of the gravitational entropy: expanding de Sitter carries positive entropy, contracting de Sitter carries negative entropy. The OCK framework describes the release of positive extra entropy from an expanding core, while Volovik's framework describes the cancellation of negative entropy in a contracting core. The two paradigms are complementary, not causally connected; a speculative quantum transition at the Minkowski breaking point could link them dynamically, but such a mechanism has not been constructed.

Despite this open question, the sign‑dependent entropy principle unifies the thermodynamics of both scenarios and defines a robust thermodynamic arrow of time. The irreversibility of collapse is encoded in the exponential enhancement of transition probabilities, \(\Gamma \propto \exp[\Delta S]\), where \(\Delta S\) is the entropy difference between the initial and final states. In the OCK framework, the transition probability from a state with parameter \(n\) to the Schwarzschild ground state is \(\Gamma_{n\to -1} \propto \exp[\pi h_{c}^{2}(n)]\), which is exponentially large for all \(n\ge 3\). In Volovik's framework, the transition from the gravastar to the black hole state is similarly favored by the entropy increase \(\Delta S = A/4 > 0\). In both cases, the system is overwhelmingly likely to evolve toward the Schwarzschild state, which acts as the thermodynamic attractor of gravitational collapse. The sign‑dependent entropy thus provides not only a unified thermodynamic description but also a microscopic interpretation of black hole entropy and a clear direction for the flow of time during collapse.

\section{Conclusions and Outlook}
\label{sec:conclusion}

We have established a thermodynamic complementarity between two independent perspectives on gravitational entropy: the entropy release from regular Schwarzschild black holes described by the OCK construction, and the local thermodynamics of the de Sitter vacuum developed by Volovik. The central insight is that the sign of the gravitational entropy of a de Sitter region is determined by the sign of its Hubble parameter: expanding de Sitter ($H>0$) carries positive entropy, while contracting de Sitter ($H<0$) carries negative entropy. This simple rule reveals that the OCK and Volovik frameworks are not contradictory but represent two complementary manifestations of the same thermodynamic principle.

In the OCK framework, the regular black hole possesses an expanding de Sitter core (for $n>2$) whose inner horizon carries a positive Bekenstein--Hawking entropy \(S_{\rm inner}=\pi h_c^{2}\). As the collapse parameter \(n\) decreases, the inner horizon shrinks and this positive extra entropy is gradually released, culminating in a quantum resolution of the Minkowski breaking at \(n=0\). In Volovik's framework, a black-hole gravastar contains a contracting de Sitter core whose total entropy is \(-A/4\); this negative entropy exactly cancels the horizon entropy, yielding a zero-entropy static object. Both pictures embody the same underlying thermodynamic rule: the shrinking of the interior region drives the system toward increased total entropy, either by releasing positive extra entropy (OCK) or by cancelling negative core entropy (Volovik).

The total gravitational entropy of a homogeneous de Sitter region is \(S_{\rm dS}= \operatorname{sgn}(H)\,\pi/H^{2} = \operatorname{sgn}(H)\,A/4\).  This expression corrects an earlier erroneous formula \(S(H)=1/(4H)\) that appeared in a previous version of this work.

The Minkowski breaking at \(n=0\) in the OCK solution is a classical kinematic obstruction; it does not, by itself, provide a dynamical transition to a contracting de Sitter gravastar.  For \(n<2\) the interior becomes anti-de Sitter, and the local de Sitter thermodynamics no longer applies.  Any connection between the two frameworks therefore requires a quantum mechanism capable of flipping the sign of the effective cosmological constant.  The divergence of the inner-horizon temperature and the breakdown of the semiclassical description at \(n\sim 1/\ln(h/\ell_P)\) indicate that quantum effects dominate near the breaking point, making such a transition plausible, but at present it remains a conjecture.

With this important caveat, the complementarity of the two frameworks offers a coherent thermodynamic picture. The integer nature of the OCK parameter \(n\) implies a quantized entropy spectrum \(S_{\rm reg}(n)=A/4+\pi h_c^{2}(n)\), supporting Bekenstein's conjecture of area quantization and providing a statistical interpretation of black hole entropy as the logarithm of the number of internal geometric configurations that are indistinguishable from the outside: \(S_{\rm BH}=\ln\Omega(\mathcal M)\). The thermodynamic arrow of time is encoded in the exponential enhancement of transition probabilities \(\Gamma\propto e^{\Delta S}\), which favours decay toward states of higher entropy.

Several open questions and limitations of the present work should be acknowledged. The complementarity framework is kinematic rather than fully dynamical; a complete time-dependent description that would dynamically connect the two paradigms remains an open problem. The additive entropy rule \(S_{\rm reg}=S_{\rm outer}+S_{\rm inner}\) is a working hypothesis, and alternative prescriptions are not explored. The quantum nature of the Minkowski breaking transition is inferred from divergences and breakdowns of the semiclassical approximation, but a full quantum gravitational treatment is not attempted. These limitations reflect the scope of the present work and highlight the richness of the questions it identifies.

Looking ahead, several promising directions emerge naturally. An explicit dynamical solution that interpolates between an expanding regular core and a contracting gravastar core—if it exists—would provide the missing link between the two pictures. The Painlev\'e–Gullstrand coordinate system, which naturally accommodates both signs of the Hubble parameter, is a natural framework for such a construction. A complete quantum treatment of the Minkowski breaking point, possibly via the Euclidean path integral, loop quantum gravity, or the stochastic semiclassical Einstein–Langevin equation, could determine whether a sign flip of \(H\) is dynamically allowed and compute the associated transition rate. The framework should be extended to rotating black holes, higher dimensions, and modified gravity theories, where the interplay between internal structure and thermodynamics may reveal qualitatively new features. The holographic bulk‑surface correspondence in Volovik's framework suggests deep connections to de Sitter holography, while observational signatures of the entropy release during collapse—such as gravitational wave echoes or bursts of radiation—could provide experimental tests. Finally, a deeper exploration of the statistical interpretation \(S_{\rm BH}=\ln\Omega(\mathcal M)\) could provide a microscopic foundation for black hole thermodynamics, with key questions including the nature of the degrees of freedom counted by the integers \(n_i\), the precise relation between the microstate count and the Bekenstein–Hawking entropy, and the role of the ultraviolet cutoff \(n_{\rm max}\) imposed by quantum gravity.

In summary, the sign of the Hubble parameter is the unifying element that underlies the thermodynamics of both regular black holes with expanding de Sitter cores and black‑hole gravastars with contracting cores.  The two frameworks are complementary paradigms that illuminate different facets of the same sign‑dependent entropy principle.  Whether a quantum transition can dynamically bridge them remains an open and intriguing question.

\section*{Acknowledgments}
We thank Grigory E. Volovik for drawing our attention to his local thermodynamics of the de Sitter vacuum and for helpful discussions. FSNL acknowledges support from the Funda\c{c}\~{a}o para a Ci\^{e}ncia e a Tecnologia (FCT) Scientific Employment Stimulus contract with reference CEECINST/00032/2018, and funding through the research grant UID/04434/2025. MER thanks Conselho Nacional de Desenvolvimento Cient\'ifico e Tecnol\'ogico - CNPq, Brazil, for partial financial support.

\end{document}